# iZone: A Location-Based Mobile Social Networking System

Rui Cheng, Zhuo Yang, Feng Xia
School of Software, Dalian University of Technology, Dalian 116620, China
f.xia@ieee.org

*Abstract*—The rapid development of wireless technology, the extensive use of mobile phones and the availability of location information are facilitating personalized location-based applications. Easy to carry, easy to use and easy to buy, smart phones with certain software are of great advantage. Consequently, mobile social networking (MSN) systems have emerged rapidly, being a revolution for our everyday life. Based on the analysis of general requirements of MSN and location-based services (LBS), this paper presents the design of iZone, a mobile social networking system, as well as a prototype implementation. This platform exploits mobile GIS (Geographic Information Systems), LBS and J2ME technologies, combining geographical data to display map on mobile phones. It can provide a number of social networking services via smartphones.

*Keywords-mobile social networking; smart phone; location-based services; mobile geographic information systems*

## I. INTRODUCTION

Social networking is a relatively new and still developing field that focuses on the analysis of social relationship. Human beings can only survive, mature, and reach self-actualization in social networking. Social networking puts its attention on social objects as, for instance, members of an association, employees of a firm, firms themselves and so on. Both theoretical advances and real-life applications are booming day by day [1, 2, 3]. Social networking exists since human beings exist. We live, work in it and may never feel it. In the history of mankind, one important technology revolution is sure to be associated with media revolution. All activities of human are essentially activities of information. During that phase time and material cost are continuously reduced by human in social networking. Cost-efficient media is always popular with human.

In 21st century, mobile phones are growing in popularity all around the world. Mobile phone has developed from a tool for chatting to a necessary friend in social networking. As it has more and more useful functions in social communication, it has gradually become an indispensable part of people's life [4]. In this case, mobile social networking (MSN) system emerges at a historic moment. Mobile social networking is a typical social network where one or more individuals of similar interests or commonalities, conversing and connecting with one another using mobile phones [5]. In the information age, it seems that people communicate using mobile devices every day.

People now do not stay in one area and move often. Cities are expanding and changing all the time, which makes people confused about the position and location. Location-based services (LBS) appear to meet people's need about location. Since Google Map started, many other companies also launched similar services to bring more convenience to people. It is not hard to see that location-based mobile social networking systems are sure to become popular.

This paper will have a glance at some popular related work all over the world in Section II, and then anatomizes MSN and LBS in Section III. We present the design of iZone, a mobile platform of location-based social networking system in Section IV. It has a number of functions supporting chatting with friends, reading the map, fixing the position, reading the news, reporting the weather and so forth. In Section V, we describe a prototype implementation of the iZone system. Section VI concludes the paper.

## II. RELATED WORK

MSN systems are developing fast around the world. There are many examples of commercialized operations. For instance, Plazes is a location-aware interaction system that helps mobile users hook up with friends or other like-minded people anywhere on the globe. Jambo Networks uses Wi-Fi-enabled laptops, mobile phones, and PDAs to match people within walking distance who have similar interests and would like to meet face to face [6]. Friendster, Dodgeball, and Facebook are three examples to promote discovery and interaction with both friends and strangers [7]. Amazon allows users to review products but no friendship among users can be defined. On the other hand, Facebook has an explicit friendship circle defined by individual users who get automatic updates on their friends' activities [8]. MSN is meanwhile a booming industry in China. China Mobile launched *Making Friends in Fetion*, which led the trend of mobile social networking in China [9].

LBS is also developing at a high speed. For instance, Bell Company in Canada is one of the market leaders of LBS. It launched services of amusement, information, calling for help based on location in December of 2003. Although My Finder is well-known around the world, it still gets rid of the stale and brings forth the fresh and offered Swordfish - a mobile game based on GPS in which it diminishes the earth into a measured fishpond. Until 2008, mobile phones supporting GPS have occupied 25 percent of all and applications are also multifarious [10, 11]. In China, LBS also has recognition in traffic area. By the end of 2009, according to incompletely statistics, scores of provinces have accomplished tracking implementation on taxi and bus, which covers vehicle position tracing, vehicle speed management and vehicle scheduling.

The importance of geographical information systems (GIS) grows with each passing day in the information-based process of national economy. Since 2000, GIS got sufficient development in the world, produced enormous economic and brought large social benefit [12]. Mobile Geographic Information System (Mobile GIS) is a product of embedded systems which has installed GIS in them. It is also an ideal solution for resource management, resource configuration, urban planning and management, land information system and cadastral application, environmental management and modeling, emergency response and so on [13, 14].

### III. BACKGROUND

#### A. Mobile Social Networking

Mobile social networking occurs in virtual communities just the same as web based social networking. Some social networking websites start to turn mobile such as MySpace and Facebook, which has become a trend many companies adopt. Meanwhile, native mobile social networks have been created like Foursquare and Gowalla. Initially, there were two basic types of mobile social networks. The first is companies that partner with wireless phone carriers to distribute their communities via the default start pages on mobile phone browsers. An example is Juice Caster. The second type is companies that do not have such carrier relationships and rely on other methods to attract users [15]. Many web based social networking companies turn mobile because more people use mobile phones to communicate and it has become a business opportunity for them. It is inconvenient for people to have computer some time and mobile phone comes into useful. Chips and cores of mobile phone are being improved in a wing-footed speed. As a result, social networking on mobile phones becomes a necessity. On the other hand, mobile social networking systems are updating quickly to meet people's needs on communication.

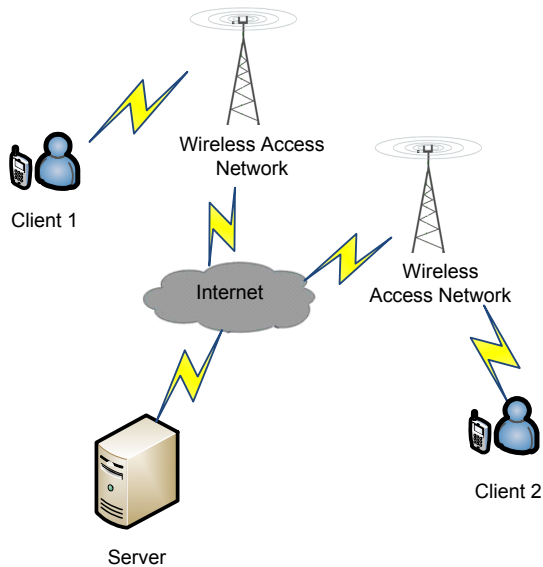

Figure 1. Structure of a MSN system

As shown in Fig. 1, a mobile social networking system generally consists of four main components: the client devices, the wireless access network, the Internet and the server side.

Client Devices must have a location module, which can receive and send data. Most mobile phones now can run Java software which has the function of location. Wi-Fi or GPS are commonly used to estimate the current location of the user. GPS requires a clear view of sky to function properly. Therefore, for the need of social services that take place in indoor environments, Wi-Fi location technology can be utilized instead of GPS [16].

The Wireless Access Network works as TCP/IP pipes. Users have to set up one basic station in a sufficient distance. By employing a proper architecture [17], changes made to the core cellular network could be minimized by e.g. removing carrier assisted positioning.

The Internet component consists of third party application servers such as SMTP Mail Server, VoIP Server, and Map Server. There are publicly available map servers that can be utilized for MSN, and that provide well-defined, albeit proprietary, APIs. The application servers in the Internet component can also be deployed in the Server Side.

The Server Side is the core of the system. It consists of Web Server, Location Database, Profile Repository, Matching Logic and Privacy Control. When the request comes, it goes through these five sections and finally sends a response to the users.

#### B. Location-Based Services

LBS systems often provide a kind of service that confirms user's actual geographical position by connection of mobile terminal and mobile network, thus meeting users' needs on information about location.

LBS can be used in a variety of aspects in our life, such as healthcare, friends making, security tracing and so forth. LBS includes services to identify the location of a person or object, such as discovering the nearest banking cash machine or the whereabouts of a friend or employee. Examples of LBS include parcel tracking and vehicle tracking services. They can include mobile commerce when taking the form of coupons or advertising directed at customers based on their current locations. They also include personalized weather services and even location-based games. These applications are typical examples of telecommunication convergence.

There are several localization methods available. The first is control plane localization. Sometimes it referred to as positioning with controlled plane. Service provider gets the location based on the radio signal delay of the closest cell-phone towers (for phones without GPS features) which can be quite slow as it uses the *voice control* channel [18]. LBS systems use a single base station, with a *radius* of inaccuracy, to determine a phone's location. This technique was the basis of the E-911 mandate and is still used to locate mobile phones as a safety measure. Latest smartphones and PDAs typically have an integrated A-GPS chip.

In order to provide a successful LBS technology the following factors must be met:

- Coordinates accuracy requirements that are determined by the relevant services;
- Lowest possible cost;
- Minimal impact on network and equipment.

Several categories of methods can be used to find the location of the subscriber. The simple and standard solution is GPS-based LBS. Sony Ericsson's *Near Me* is one such example. It is used to maintain knowledge of the exact location. However, this can be expensive for the end-users, as they would have to invest in a GPS-equipped handset. GPS is based on the concept of trilateration, a basic geometric principle that allows finding one location if one knows its distance from other already known locations.

GSM localization is the second option. Finding the location of a mobile device in relation to its cell site is another way to find out the location of an object or a person. It relies on various means of multilateration of the signal from cell sites serving a mobile phone [19]. The geographical position of the device is found out through various techniques like time difference of arrival (TDOA) or Enhanced Observed Time Difference (E-OTD).

Another method is Near LBS (NLBS), in which local-range technologies such as Bluetooth, WLAN, infrared and/or RFID/Near Field Communication technologies are used to match devices to nearby services [20]. This technique allows a person to access information based on their surroundings, especially suitable for using inside closed premises, restricted/ regional areas.

Another alternative is an operator- and GPS-independent location service based on access into the deep level telecoms network (SS7) [21]. This solution enables accurate and quick determination of geographical coordinates of mobile phone numbers by providing operator-independent location data and works also for handsets that are not GPS-enabled.

Many other local positioning systems are available, especially for indoor use. GPS and GSM don't work very well indoor, where other techniques are often used, including Bluetooth, UWB, RFID and Wi-Fi.

## IV. iZone System Design

iZone is designed to be a comprehensive, secure, convenient and easy-handling platform. It is a J2ME project that can run on mobile devices which support CLDC 1.0 and MIDP 2.0. This platform is a type of intelligent location-based system which combines GIS and location information. iZone can collect position data and other space information, then automatically extract things that may be interested by users, and finally send them to the users in a humanizing form. Meanwhile, it offers users some social networking services, such as instant messaging, blog publication, photo uploading and position tracing based on users' location.

As shown in Fig. 2, iZone is mainly divided into the following seven subsystems: Registration and Login Subsystem, Map Subsystem, Homepage Subsystem, Friend Subsystem, Mail Subsystem, Local Information Subsystem and Setting Subsystem.

Each subsystem can also be divided into several modules. Registration and Login Subsystem consists of login module and registration module. Map Subsystem consists of local search module and on-line friends module. Homepage Subsystem consists of friend news module, gateway module, personal information module, blog module and album module. Friend Subsystem consists of friend management module and friend communication module. Mail Subsystem is an individual system. Local Information Subsystem consists of local weather report module, local news module and local forum module. Setting Subsystem consists of privacy setting module and information setting module.

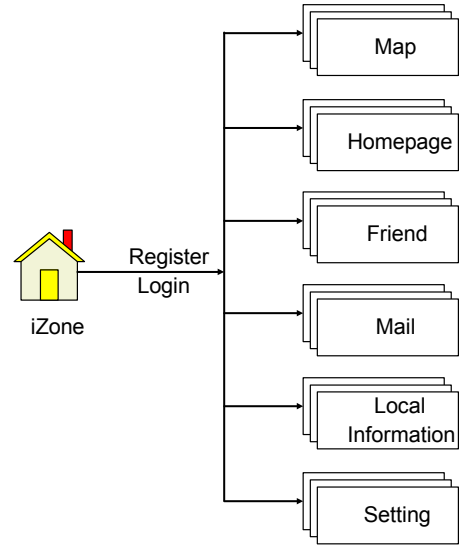

Figure 2. iZone system structure

### A. Registration and Login Subsystem

Registration and Login Subsystem is the first subsystem the user will meet when using this software. It consists of registration and login modules. In Registration module, users must fill in some basic information such as unique user name, password, nickname, E-mail, phone number and gender. When the registration apply was authenticated, the server will send a short message to user's mobile phone and then the user can login. Only by inputting correct user name and password, can the user login successfully.

### B. Map Subsystem

After login, the user can see the default subsystem, Map Subsystem, as shown in Fig. 3. It is the core of location based system. It consists of local search and on-line friends modules. In local search module, the user can search local users, local restaurants, and local hospitals and so on. It is a lively guide for a visitor. User can also choose typical view and satellite view. In on-line friends module, user can see other nearby users on the map. Some further information (e.g. nickname, gender, and status) of a certain user will be shown if his/her avatar is pointed at. Moreover, the map has the function of Zoom In and Zoom Out. It is easy to change the map size.

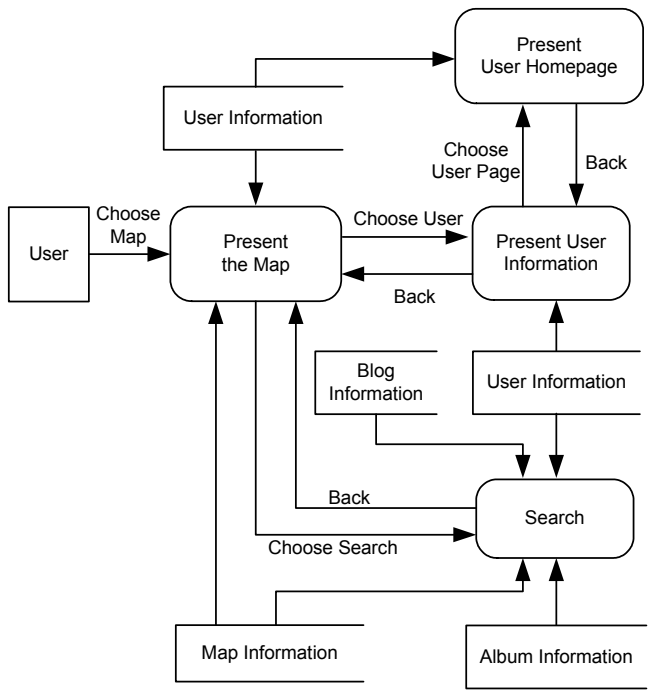

Figure 3. Dataflow diagram of Map Subsystem

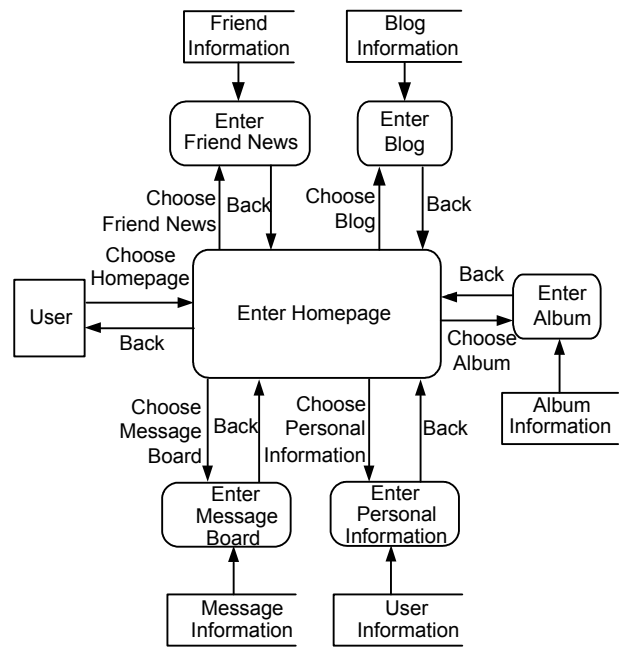

Figure 4. Dataflow diagram of Homepage Subsystem

*C. Homepage Subsystem*

Homepage Subsystem is a collection of friend news, gateway, personal information, log and album modules, as shown in Fig. 4. In friend news module, all the dynamic information about friends can show up here, such as Alice changed her avatar, Bob published a new blog, Catherine uploaded many new photos and so on. All these news will be listed according to the corresponding time points. New information will appear in front of old ones and user can comment on this news.

Homepage module is the door to other modules. It mainly records the user's trace, others who has visited this page and the user's friend lists. By clicking on the avatar of one of his/her friends, user can go to visit the page. Personal information module mainly consists of four sections of information: basic information, contact information, location information and privacy information. After login, user can edit all these information so that making friends with similar information will be easier. In blog module, it is available to publish, edit, delete, view and comment on the blogs. User can write his own blog and choose to save it in the draft or publish it. What's more, user can manage his blog and make comments on other users' blog. Album module is the last module of homepage subsystem. It consists of collection management and photo management. User can put some photos in one photo album and it will be easier to manage them. It is also available to upload, edit, delete and comment on them like blog. What may be different is that user has to upload photos from mobile phone to system.

*D. Friend Subsystem*

Friend Subsystem is sure to popular in the youth. It consists of friend management and friend communication modules. In friend management module, user can set up groups and put different friends into different groups. User can also add, edit, delete and view groups. For the first time a user enter this system, he/she has two default groups: My Friends and Strangers. User can also add, delete, view and edit friends. There is one function named friend recommendation, in which you can make friends with similar interests. In friend communication module, user can chat with each other like other instant message software. User can also transfer pictures and files. By default, user' communication records will be saved. The user can of course change the setting.

*E. Mail Subsystem*

Mail Subsystem is a single system with the function of dealing with mails. User can write, send, view, delete and manage mails. It is similar to common mail system but a little downsized. When a new letter comes, it will show in internal letter to inform user.

*F. Local Information Subsystem*

Local Information Subsystem can give user news about local city. It consists of local weather report, local news and local forum modules. In local weather report module, user can get weather forecast about next three days including temperature, sunshine, humidity and wind velocity. It also offers some suggestions accordingly such as remember to bring umbrella or take more clothes. Local news module tells user news every day. User can subscribe different sections of news like sports, health or politics etc. Local forum module is the forum for users to post messages. User can discuss based on politics, social phenomenon and so on. User can also respond others in the forum. There are also some administrators to examine and verify messages.

*G. Setting Subsystem*

Setting Subsystem is the system for user to set personalized information and privacy. It consists of privacy setting module and information setting module. In privacy setting module, user can set whether the others are allowed to view his information such as mobile phone number, gender, birthday etc. User can set this as *all people are allowed*, *only friends are allowed* or *nobody else is allowed*. In information setting, user can edit his/her personal information. It is the shortcut of personal information module of Homepage Subsystem. User can edit them here without entering Homepage Subsystem.

V. PROTOTYPE IMPLEMENTATION

The proposed system architecture presents a location-based system named iZone based on a combination of wireelss and J2ME technologies for identifying the location of user and provides a platform for users to communicate with each other. To this moment we have finished the development of a prototype system of iZone. This section will introduce this prototype implementation.

*A. Registration and Login*

iZone can work on mobile phones which support CLDC 1.0 and MIDP 2.0. After installing it on mobile phone, a user can start it and it is necessary to register for the first time.

First of all, fill in username, which must be unique in the system; otherwise the screen will pop up a warning window and ask the user to fill another one. It is also required to fill in password, nickname, e-mail address and phone number. If the username being input has been occupied by others, a window with a warning message will appear to ask the user to change another one. After finishing them, just press *Submit* button. All the information will be sent to the database server.

The user has to input correct username and password and then can enter the system. If the user input wrong username or password, the system will ask if you have forgotten your password. Then the user can press yes and get back your password through a series of certification.

If the mobile phone has access to wireless Internet and correct information has been input, a wonderful tour in iZone after a few seconds of authentication will bring to the user.

*B. Localization*

After successful login, the Map Subsystem appears automatically. iZone first needs to get the position of the mobile phone and the map type the user want to see. For example, the user inputs Dalian, China and chooses normal map view. The user has to input city name first, a comma behind and country name at last. There are three kinds of map type: normal, satellite and hybrid. In normal view of map system, user can view the city like reading maps. Important railways and highways can be seen on it. In satellite view of map subsystem, user can view the city on satellite images. In hybrid view of map subsystem, as it is called, user can see both kinds of images, as shown in Fig. 5.

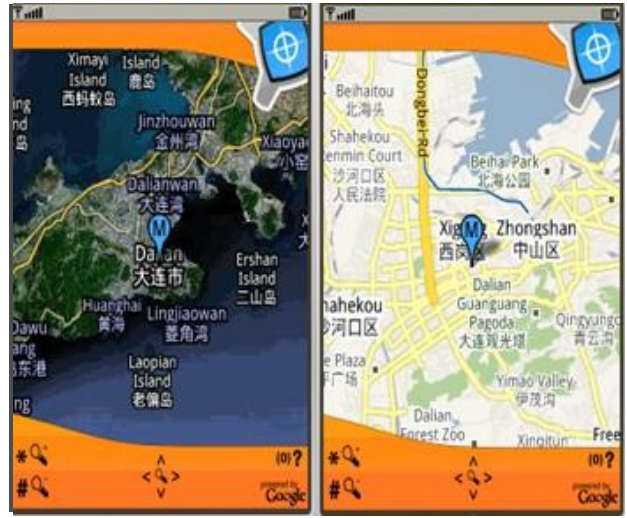

Figure 5. Localization interfaces

Lots of functions are available in this subsystem. User can press button 1 on mobile phone to switch to normal view; press 2 to satellite view and 3 to hybrid view. If the user wants to see the map clearly, just press the * button to zoom in the map and # to zoom out it.

*C. Friend List and Chatting*

After enjoying Map Subsystem, user must be eager to talking with friends via this platform. The user can enter Friend Subsystem to chat with friends. The friends will be listed automatically on the screen, as shown in Fig. 6. On-line friends are shown in full-color mode and off-line friends are shown in gray-color mode. User can see avatars of users in front of their names. User can also see their usernames in the brackets.

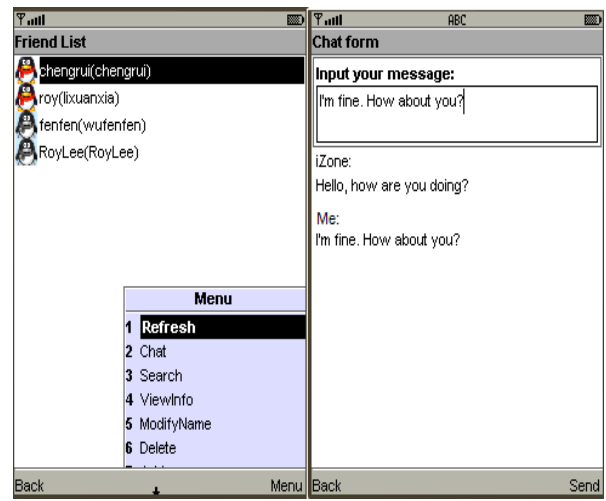

Figure 6. Friend list and chatting interface

There is a *refresh* button. If it is pressed, the screen will be updated so that the newest friend list will appear. User can choose any friend and chat with him/her. If user wants to find new friends, he/she could choose *search* and find friends by username, city or interests. If the user wants to find a friend's birthday, just choose view information and it will appear on the screen. Moreover, the user can also change friends' nickname and remove friends.

## VI. CONCLUSIONS

This paper has presented the design and a prototype implementation of our iZone system. The system is based on a combination of wireless technology, J2ME, LBS and GIS and is able to locate users and send information based on their locations. A powerful iZone system is still under implementation. We believe that the wide spread of mobile phones and wireless technology will lead to extensive development of location-based applications.

## ACKNOWLEDGMENT

The authors thank Xuanxia Li, Aili Dong, Fenfen Wu, and Han Liu at Dalian University of Technology for their help in programming. This work was partially supported by the National Natural Science Foundation of China under Grant No. 60903153 and the Fundamental Research Funds for the Central Universities.

## REFERENCES

[1] Wei Wang, Hong Man and Yu Liu, "An Intrusion Detection System in Ad Hoc Networks: A Social Network Analysis Approach," 6th IEEE Consumer Communications and Networking Conference (CCNC), Jan. 2009, pp. 1-5.

[2] Juwel Rana, Johan Kristiansson, Josef Hallberg and Kare Synnes, "An Architecture for Mobile Social Networking Applications," First International Conference on Computational Intelligence, Communication Systems and Networks (CICSYN), July. 2009, pp. 241-246.

[3] Matteo Brunelli and Michele Fedrizzi, "A fuzzy approach to social network analysis," International Conference on Advances in Social Network Analysis and Mining (ASONAM), July. 2009, pp. 225-230.

[4] Carolyn Wei and Beth E. Kolko, "Studying Mobile Phone Use in Context: Cultural, Political, and Economic Dimensions of Mobile Phone Use," International Professional Communication Conference (IPCC), July. 2005, pp. 205-212

[5] G.G. Marta, A.H. Cesar, and B. Albert-Laszlo, "Understanding individual human mobility patterns," Nature, Vol. 453, No. 7196, 5 June 2008, pp. 779-782.

[6] Yao-Jen Chang, Hung-Huan Liu, Li-Der Chou, Yen-Wen Chen and Haw-Yun Shin, "A General Architecture of Mobile Social Network Services," International Conference on Convergence Information Technology, Nov. 2007, pp. 151-156.

[7] Soulakshmee D. Nagowah, "Aiding Social Interaction via a Mobile Peer to Peer Network," Fourth International Conference on Digital Society, Feb. 2010, pp. 130-135.

[8] Guanling Chen and Faruq Rahman, "Analyzing Privacy Designs of Mobile Social Networking Applications," IEEE/IFIP International Conference on Embedded and Ubiquitous Computing, Dec. 2008, pp. 83-88.

[9] Weimin Wang, Liangliang Liu, Dongsheng Wang, Yanan Cao, Yuming Wu, Baoyuan Qi, Dongchun Qiao, Jingyang Guo, Yufei Zheng, Cungen Cao, Li Jianghua, Shi Yuan, Li Ruilin, Sang Changqin and Shen Nan, "An Intelligent Call Center Platform for Mobile Communication Enterprises," International Conference on Communications and Mobile Computing (CMC), April. 2010, pp. 370-375.

[10] Basit Qureshi, Geyong Min, Demetres Kouvatsos and Mohammad Ilyas, "An Adaptive Content Sharing Protocol for P2P Mobile Social Networks," IEEE 24th International Conference on Advanced Information Networking and Applications Workshops (WAINA), April. 2010, pp. 413-418.

[11] Anthony Di Michele and Jean Huppe, "Evolving a Fibre-to-the-Node Access Infrastructure," Optical Fiber Communication Conference and National Fiber Optic Engineers Conference, March.2006, pp. 6.

[12] Feixiang Chen, Chongjun Yang, Wenyang Yu, Xiaoqiu Le and Jianyu Yang, "Research on Mobile GIS Based on LBS," IEEE International Geoscience and Remote Sensing Symposium, July. 2005, pp. 4.

[13] Shu-liang Zhang, Shao-song Ma and Yi-ming Zhang, "Research on Collaborative Environment of Data Collection and Application in Mobile GIS," Third International Conference on Multimedia and Ubiquitous Engineering, June. 2009, pp. 421-428.

[14] Qinghui Sun, Xiaoli Wang, Tianhe Chi and Cuiling Ji, "An integrated system based on wireless communication technology and mobile GIS," IEEE International Geoscience and Remote Sensing Symposium, July. 2005, pp. 963-966.

[15] Gill Clough, "Geolearners: Location-Based Informal Learning with Mobile and Social Technologies," IEEE Transactions on Learning Technologies, March. 2010, pp. 33-44.

[16] L.-D. Chou and C.-Y. Chang, "A hierarchical architecture for indoor positioning services," Joint Positioning, Navigation and Communication, Series in Hannoversche Beiträge zur Nachrichtentechnik, Editor: Kyamakya and Kyandoghere, Shaker Verlag Publishers, Germany, ISBN: 3832237461, Mar. 2005.

[17] S. J. Barbeau, M. A. Labrador, P. L. Winters, R. Perez, and N. L. Georggi, "A General Architecture in Support of Interactive, Multimedia, Location-Based Mobile Applications," IEEE Communications Magazine, Nov.2006, pp. 156-163.

[18] Balqies Sadoun and Omar Al-Bayari, "LBS and GIS Technology Combination and Applications," IEEE/ACS International Conference on Computer Systems and Applications, May.2007, pp. 578-583.

[19] Wei Li, Suxiang Li, Yunchun Li, Shasha Luo and Xinwei Chen, "An in-door wireless localization system oncampus network," 4th International Conference on Wireless Communications, Networking and Mobile Computing, Oct. 2008, pp. 1-4.

[20] P. Castro, P. Chiu, T.Kremenek, and R.A. Muntz, "Probabilistic Location Service for Wireless Network Environments," Int Conf. on Ubiquitous Computing (Ubicomp), September 2001.

[21] Paramvir Bahlnd and Venkata N. Padmanabhan, "A Software System for Locating Mobile Users: Design, Evaluation, and Lessons," Microsoft Research, Anand Balachandran, 2000.